\documentclass[epj-spec]{svjour}
\usepackage{graphics}
\begin{document}
\title{A cosmological model for corrugated graphene sheets.}
\author{Alberto Cortijo
\inst{1}
\fnmsep
\thanks{\email{cortijo@icmm.csic.es}}
\and Mar\'ia A. H. Vozmediano\inst{2}  }
\institute{Instituto de Ciencia de Materiales de Madrid,
CSIC, Cantoblanco, E-28049 Madrid, Spain.\and Universidad Carlos III de Madrid,
Avenida de la Universidad 30, E-28913 Legan\'es, Madrid, Spain}
\abstract{
Defects play a key role in the electronic structure of graphene layers flat or curved.
Topological defects in which an hexagon is replaced by an n-sided polygon generate
long range interactions that make them different from vacancies or other potential defects.
In this work we review previous models for topological defects in graphene. A formalism
is proposed to study the electronic and transport properties of graphene sheets with
corrugations as the one recently synthesized. The formalism is based on coupling the
Dirac equation that models the low energy electronic excitations of clean flat
graphene samples to a curved space. A cosmic string analogy allows to treat an
arbitrary number of topological defects located at arbitrary positions on the
graphene plane. The usual defects that will always be present in any graphene
sample as pentagon-heptagon pairs and Stone-Wales defects are studied as an example.
The local density of states around the defects acquires characteristic
modulations that could be observed in scanning tunnel and transmission
electron microscopy.
} 
\maketitle
\section{Introduction.}
\label{intro}
The recent synthesis of single layers of graphite -graphene -
\cite{Netal05,Zetal05}, has opened the
way to study a new really two-dimensional system that constitutes an excellent
laboratory to test some of the most fruitful ideas of recent condensed matter:
localization, existence-or not- of Fermi liquids in two dimensions, or the
nature of the metallic state in two dimensions. Moreover the the experiments
have been able to verify the beautiful theoretical model predicted years
ago for the low energy excitations described by the massless Dirac
equation\cite{theoryold} in two dimensions \cite{theorynew,GGV92,GGV93} and
permit to envisage graphene as a laboratory to test field theory or
cosmological ideas\cite{KNG06}.

Most of the experimental samples studied in recent times show
mesoscopic corrugations\cite{G06} which are produced in the
process of cleavage and are
observed by atomic force microscopy.
The height of the observed  ripples go up to
several Amstrongs and the lateral size is typically a few tens of
nanometers.
They have been invoked
to explain the reported absence of weak (anti)-localization observed
in graphene\cite{Metal06,MG06}
and  to strengthen the effects of the
spin orbit coupling \cite{HGB06}.
The presence of this curved regions in graphene can have strong effects
on  the physical properties and has not yet been fully explored.

A systematic study of the electronic properties of slightly curved graphene
sheets has been started in \cite{CV06a,CV06b} where it is suggested
that the observed ripples are due to the presence of heptagon-pentagon pairs or
Stone-Wales defects in the graphene sheet. This type of defects have
been observed in the previously existing forms of graphene (nanotubes
and fullerenes) and should be very natural in the flat samples.
In this work we explain the derivation  of a  model based on a
cosmological analogy that can describe an average flat
graphene sheet with regions of non zero curvature
induced by an arbitrary number of heptagons and pentagons. We also
present some results on the charge inhomogeneities to be expected
in samples with this type of disorder.

\section{Disorder in graphene.
Experimental observation of topological disorder in graphene
and related compounds.}
\label{sec:1}
Disorder has a very strong influence on the electronic properties of
low dimensional systems. During the "first revolution" of graphene after
the synthesis of fullerenes and nanotubes, there appeared many works treating
vacancies, adatoms, edges, and what we call topological disorder produced
by substitution of a hexagonal ring in the honeycomb of graphene by an
n-sided polygon.  Observations of all these types of disorder were
reported in \cite{Hetal04}. Topological disorder was responsible
for the formation of fullerenes and had also a strong
influence on the electronic properties of nanotubes. One of
the most popular types of defects in nanotubes were the
Stone-Wales defects made by a $\pi/2$ rotation of a graphene bond giving
rise to the formation of two pentagons and to heptagons.
\begin{figure}
  \begin{center}
  \resizebox{0.75\columnwidth}{!}{%
   \includegraphics{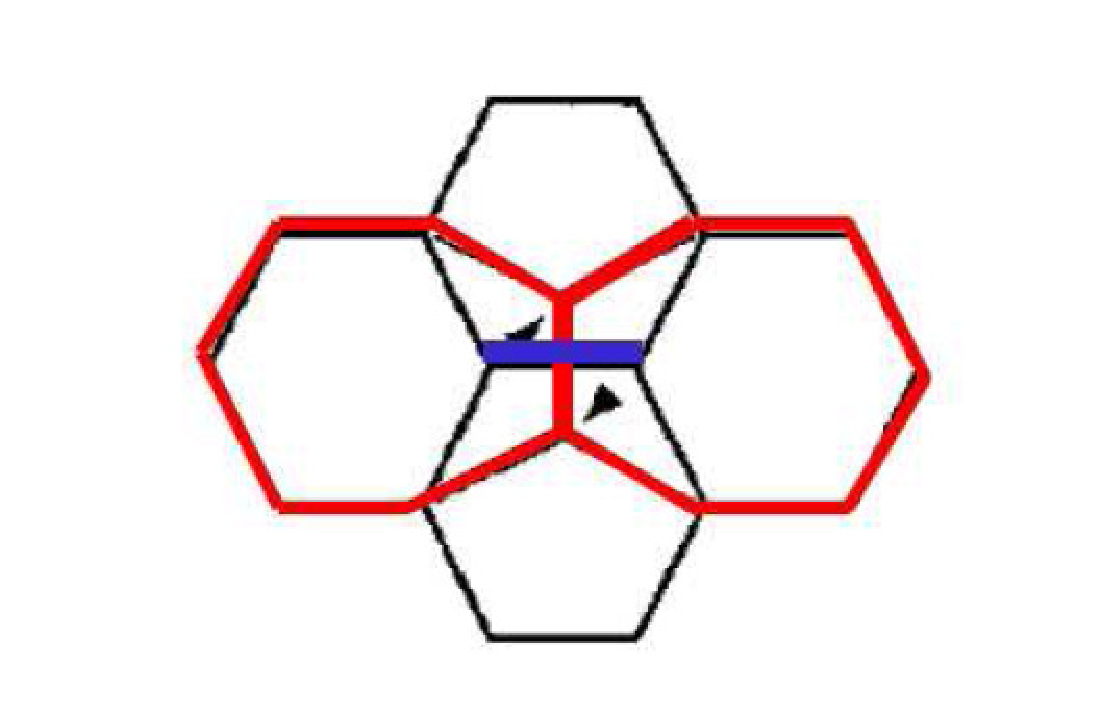} }
    \caption{Formation of a Stone Wales defect in the honeycomb lattice.
    }
    \label{SW}
\end{center}
\end{figure}
Among the topological defects these have minimal  activation energy
and their presence allows to design   nanotubes which are metallic at
one end and semiconducting at the other. Pentagon-heptagon
pairs were also used to join nanotubes of different chiralities
or diameters \cite{Cetal96}. A very important property of the
topological defects that will be discussed in this work is
that they generate a long range potential. This makes a big difference
with vacancies or impurities modelled by local potentials.

Topological defects can be seen as disclinations of
the lattice which acquires locally a finite curvature. The accumulations of various
defects may lead to closed shapes. Rings with $n<6$ sides give rise to positively
curved structures, the most popular being the $C_{60}$ molecule that has twelve pentagons.
Polygons with $n>6$ sides lead to negative curvature as occur at the joining
part of carbon nanotubes of different radius and in the Schwarzite\cite{Petal03},
a structure appearing in many forms of carbon nanofoam\cite{Retal04}.
This type of defects have been observed in experiments
with carbon nanoparticles\cite{Hetal04,Getal94,Ketal97,Detal92}
and other layered materials\cite{TTM01}.
Conical defects with an arbitrary opening angle
can be produced by accumulation of pentagons in the
cone tip and have been observed in \cite{JRDG03,Anetal01}.
Inclusion of an equal number of pentagons and heptagonal rings in a graphene
sheet would keep the flatness of the sheet at large scales and produce
a flat structure with curved portions that would be structurally stable
and have distinct electronic properties.

\section{Modelling topological disorder. }
\label{sec:2}
 As it is well known, the low-energy excitations with momenta
in the vicinity of any of the Fermi points $K_+$ and $K_-$ of graphene
have a linear dispersion
and can be described by a continuous model which reduces to the
Dirac equation in two dimensions\cite{theoryold}.
In the absence of interactions or disorder mixing  the two
Fermi points, the electronic properties of the system around
the Fermi point $i$ are well described
by the effective low-energy Hamiltonian density:
\begin{equation}
{\cal H}_{0i}= \hbar v_{\rm F}  \bar{\Psi}_i({\bf r})
(  \sigma_x \partial_x + \sigma_y \partial_y )
\Psi_i ({\bf r})\;,\label{hamil}
\end{equation}
where $\sigma_{x,y}$ are the Pauli matrices,
$v_{\rm F} = (3 t a )/2 $, and $a=  1.4\AA$ is the
distance between nearest carbon atoms. The
components of the two-dimensional wavefunction:
\begin{equation}
\Psi_i( {\bf r} )= \left( \begin{array}{c}
\varphi_A ( {\bf r} ) \\  \varphi_B ( {\bf r }) \end{array} \right)
\label{2spinor}
\end{equation}
correspond to the amplitude of the wave function in each of the two
sublattices (A and B) which build up the honeycomb structure.

\begin{figure}
  \begin{center}
  \resizebox{0.75\columnwidth}{!}{%
   \includegraphics{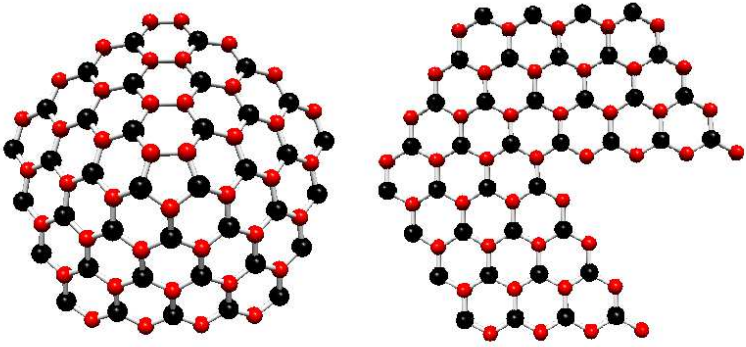} }
    \caption{Left: Effect of a pentagonal defect in a graphene layer.
    Right: Cut-and-paste procedure to form the pentagonal defect. The  points at the
    edges are connected by a link what induces a frustration of the bipartite character of
    the lattice at the seam.
    }
    \label{fig1}
\end{center}
\end{figure}

Substitution of an hexagon by an n-sided
polygon in the graphene lattice can be described by a cut-and-paste procedure
as the one shown in fig. \ref{fig1}  for the particular case of a pentagon.
A $\pi/3$ sector of the lattice is removed and the edges are glued.
In this case the planar lattice acquires the form of a cone with the pentagon in
its apex.
Such a disclination has two distinct effects on the
graphene sheet. It induces locally positive (negative) curvature
for $n<6$ ($n>6$) and, in the paste procedure, it
can break the bipartite nature of the lattice if n is odd
while preserving the symmetry if n is even. This makes a difference
with the case of the formation of nanotubes where the bipartite
nature of the lattice always remains intact.
There are then two distinct effects to take into account: the curvature
that we will treat later, and the non-trivial boundary condition that
the frustration induces on the spinor field. Both effects are well known
in other branches of physics, the second is related to the holonomy or
Berry phase.

In relativistic quantum field theory where there is a very tight connection
between the spin  and the statistics, a spin 1/2 particle is described by a field
that belongs to the so-called spinor representation of the Lorentz group.
A distinct characteristic of spinor representations is that when they
move around a closed path they pick up a minus sign (upon a $2\pi$
rotation they acquire a phase of $\pi$). A spinor going around a closed path
encircling the pentagon in Fig. 1 will acquire a phase proportional to one half
the total angle of the path ($2\pi-\pi/3=2\pi(1-1/6))$. When solving
the Shrodinger equation $H\Psi = E\Psi$ with the hamiltonian
(\ref{hamil}) this condition has to be imposed on the wave function
as a boundary condition which often are hard to deal with. A way
to incorporate the constraint attached to the boundary condition to
the hamiltonian  is to
remember the Bohm-Aharonov (BA) effect and substitute the pentagon by a
fictitious magnetic field located at the same point. The flux of the field
can be adjusted so that the phase acquired by the spinor when going around
the vector potential is the same as the one induced by going around the pentagon.
 This procedure has the advantage to be easily generalized to defects of
 arbitrary opening angle. In
the BA effect the phase is proportional to the circulation of
the vector potential along the closed path:
\begin{equation}
\oint {\bf A.dr}=2\pi(1-\pi/3).
\label{vortex}
\end{equation}
 The simplest vector potential having the property (\ref{vortex})
 is a vortex:
 $${\bf A}(x,y)=(\frac{-y}{x^2+y^2},\frac{x}{x^2+y^2})
 =-{\bf \nabla} \theta(r),$$ which in polar coordinates reduces to the
 gradient of the polar angle $\theta$.
The presence of a pentagon in the lattice has an additional
consequence which was discussed at length in \cite{GGV93}:
by going along a closed path encircling the pentagon
the two Fermi points are also exchanged. This new
condition  can be  fulfilled by attaching a quantum number
(flavor) to the Fermi points, and considering the two bispinors
$\psi_i({\bf r}), i=1,2$ of eq. (\ref{2spinor}) as
 the two components of an SU(2)
flavor doublet. The vector field will now be a non-abelian gauge
field able to rotate the spinors in this flavor space. The full
boundary condition to impose on the spinor when circling a
pentagon (or any conical singularity of arbitrary defect angle $\varphi$) is
\begin{equation}
\Psi(\theta=0)=T_{C}\Psi(\theta=2\pi)\Leftrightarrow\Psi(\theta=0)=
\exp({\oint_{_{C}}\textbf{A}_{a}T^{a}d\textbf{r}})\Psi(\theta=2\pi),
\label{boundary}
\end{equation}
$$\oint {\bf A.dr}=2\pi-\varphi.$$
where $\textbf{A}_{a}$ are a set of gauge fields and $T^{a}$ a set
of matrices related to the flavor SU(2) degree of freedom of the
system.
The situation becomes  more complicated when there are various conical
defects in the sample.  The problem of the holonomy. i.e. of the
transformation properties of the spinor upon parallel transport around
closed paths ca be generalized as described in \cite{LC04,CV06c}, the
effect of the curvature is much harder to treat.

 In \cite{GGV92} the gauge model was used to compute the electronic
 spectrum of the $C_{60}$ and other spherical fullerenes. To account for
 curvature effects, we solved the Dirac equation
 on the surface of  a sphere -thereby smoothing the curvature singularities-.
The effect of the fictitious magnetic field  was
in turn smoothed over the sphere by considering an
abelian  monopole sitting at its center
with the total magnetic charge appropriate to account for the 12 pentagons
needed to close the structure.
The simplified model for the icosahedral C60 with an abelian monopole
of charge 3/2 reproduced the observed low-energy spectrum and
the comparison with more detailed
calculations was quite reasonable.

\section{A cosmological model.}

The approach described in the previous section allows to account for both
the curvature and the holonomy induced by topological defects  when
the graphene sheet is wrapped on a surface with an explicit parametrization. It
has been applied to the case of the ellipsoid or the hiperboloid in\cite{OKP05}.
The system that we try to describe, namely, an asymptotically flat graphene sheet
with some portions curved, can be modelled by the presence of
an equal number of pentagons and heptagons in the lattice. Modelling
the heptagons is a problem and so it is to include the curvature.
Conical singularities have been studied
in cosmology as they are produced by  cosmic strings, a type of
topological defect that arises when a U(1) gauge symmetry is
spontaneously broken\cite{Vilenkin}.
The authors of ref. \cite{AHO97} study the Green's function
of different fields in  a cosmological scenario with an arbitrary
number of cosmic strings.
The metric of a two dimensional space in presence of a single cosmic
string in polar coordinates is:
\begin{equation}
ds^{2}=-dt^{2}+dr^{2}+c^{2}r^{2}d\theta^{2},\label{metric}
\end{equation}
where the parameter $c$ is a constant related to the deficit angle b
(=$2\pi/6$ in the case of a pentagon)
by $c=1-b$.
The case of   a single cosmic string which
represents a deficit angle in the space can be generalized
to describe seven membered rings representing an angle
surplus by considering a value for c
larger than $1$.  This situation is non-physical from a general
relativity viewpoint as it would correspond to a string with negative mass
density but it makes perfect sense in our case. The scenario can also
be generalized to describe an arbitrary number of pentagons and
heptagons by using  the following metric:
\begin{equation}
ds^{2}=-dt^{2}+e^{-2\Lambda(x,y)}(dx^{2}+dy^{2}),\label{genmetric}
\end{equation}
where $$\Lambda(\textbf{r})=\sum^{N}_{i=1}4\mu_{i}\log(r_{i})$$ and
$$r_{i}=[(x-a_{i})^{2}+(y-b_{i})^{2}]^{1/2}.$$ This metric describes
the space-time around N parallel cosmic strings, located at the
points $(a_{i},b_{i})$. The parameters $\mu_{i}$ are related to the
angle defect or surplus by the relationship $c_{i}=1-4\mu_{i}$ in
such manner that if $c_{i}<1 (>1)$ then $\mu_{i}>0 (<0)$.

The dynamics of a massless Dirac spinor in a curved spacetime is
governed by the Dirac equation:
\begin{equation}
i\gamma^{\mu}\nabla_{\mu}\psi=0 \label{dircurv}
\end{equation}
The difference with the flat space  lies in the definition
of the $\gamma$ matrices that satisfy generalized anticommutation
relations
\begin{equation}
\{\gamma^{\mu},\gamma^{\nu}\}=2g^{\mu\nu},\nonumber
\end{equation}
and in the covariant derivative operator, defined as
$$
\nabla_{\mu}=\partial_{\mu}-\Gamma_{\mu}
$$
where $\Gamma_{\mu}$ is the spin connection of the
spinor field that can be
calculated using the tetrad formalism\cite{birrell}.
From equation (\ref{dircurv}) we can
write down
the Dirac equation for the electron propagator, $S_{F}(x,x')$:
\begin{equation}
i\gamma^{\mu}({\bf r})(\partial_{\mu}-\Gamma_{\mu})S_{F}(x,x')=
\frac{1}{\sqrt{-g}}\;\delta^{3}(x-x'),
\label{propcurv}
\end{equation}
where $x=(t, {\bf r})$.
The local density of states $N(\omega,\textbf{r})$ is obtained from (\ref{propcurv})
by Fourier transforming the time component and taking the limit
${\bf r}\to {\bf r'}$:
\begin{equation}
N(\omega,\textbf{r})=Im
Tr S_{F}(\omega,\textbf{r},\textbf{r}).\label{LDOS}
\end{equation}
We solve eq. (\ref{LDOS}) considering the  curvature induced
by the defects as a perturbation of the flat graphene layer.

\section{Effect on the density of states.}
\begin{figure}
  \begin{center}
  \resizebox{0.75\columnwidth}{!}{%
   \includegraphics{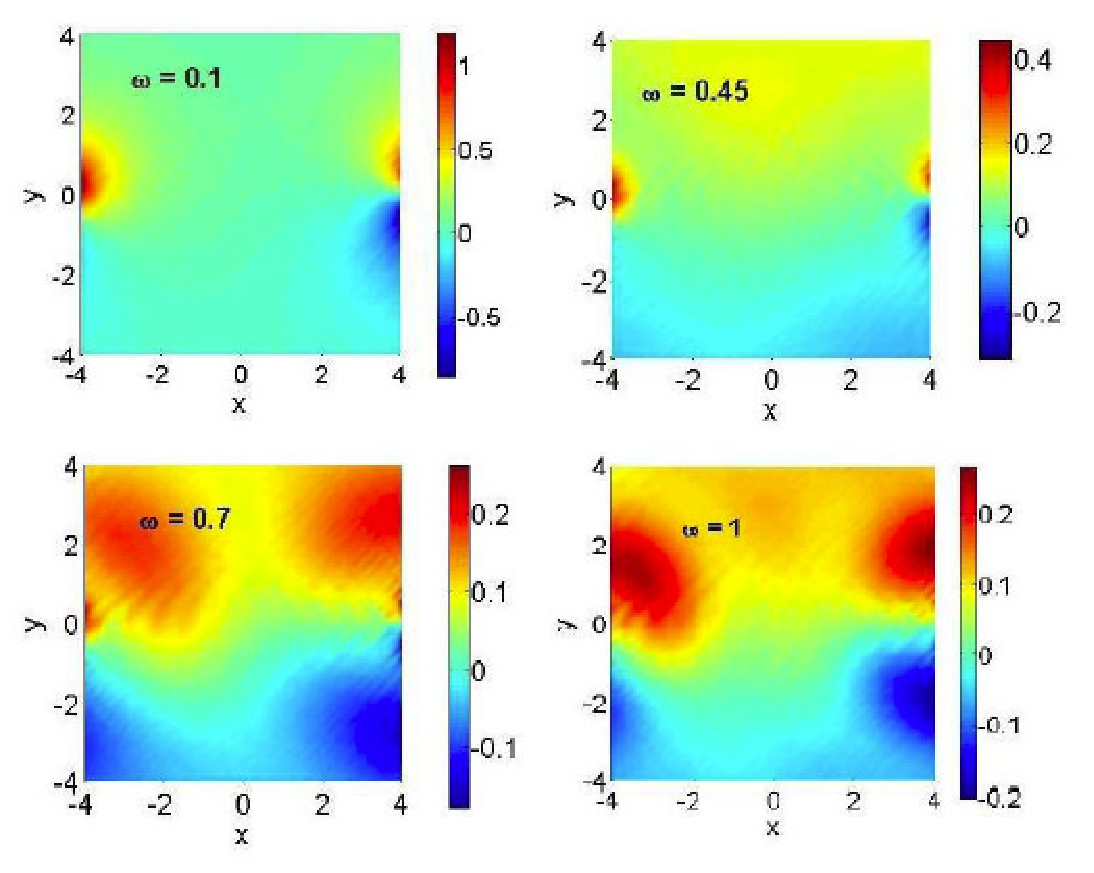} }
    \caption{Correction to the    local density of states in a wide region around two pairs
    of heptagon-pentagon defects located out of the region (see text)
    for increasing values of the energy.
    }
    \label{composition}
\end{center}
\end{figure}
With the model presented in the previous section we are able to compute the
density of states of a system with "ripples". The details of the
calculation are given in \cite{CV06b}. The main results are that
the local density of states is enhanced around defects with $n<6$ which induce
positive curvature in the lattice while the charge is
"repelled" from regions with negative curvature ($n>6$).
Heptagon-pentagon
pairs that keep the graphene sheet flat
in the long range behave as dipoles
and give rise to characteristic modulations of the DOS that can be observed
by Scanning Tunnel Microscopy. The intensity of the oscillations grows with the energy.
Fig. \ref{composition} shows the relative correction (normalized to the
free density of states) to the local density of states in a
extended region of the lattice
induced by two pairs of heptagon-pentagon defects located out of the region
for increasing values of the energy. The color code is indicated in
the figure: green stands for the DOS of perfect graphene at the given energy
and red (blue) indicates an accumulation (depletion) of the density in the
area. The patterns depend also on the relative orientation of the dipoles.
The spacial extent of the correction is such that the intensity
decays to ten percent in approximately 20 unit cells so they
can be observed in  scanning tunnel spectroscopy as
inhomogeneous regions of a few nanometers.

Another example  is despicted in fig. \ref{SW} which shows
the structure of the local density of states in real space induced in the
graphene sheet by the presence of an Stone Wales defect.
The left side shows the corrections due to  a Stone-Wales defect located in the middle
of the area
at a fixed intermediate frequency. The modulation of the local
density of states around the defects is hardly noticeable due to the strong
intensity localized at the defects. The same image is shown in the right hand side of
fig. \ref{SW} with the Stone-Wales defect located out of
plane in the upper right corner. The modulation of the LDOS is clearly
visible. The correction obtained is of the order of a few percent.
\begin{figure}
  \begin{center}
  \resizebox{0.75\columnwidth}{!}{%
   \includegraphics{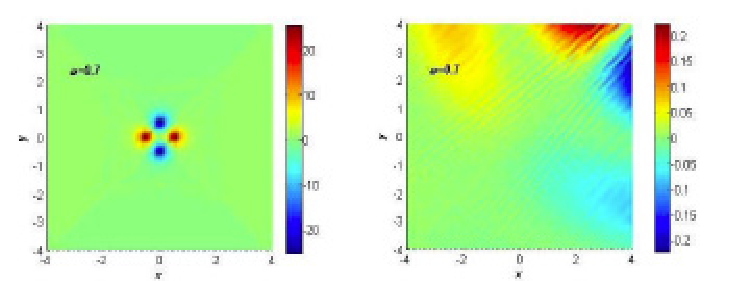} }
    \caption{Local density of states around a Stone-Wales defect
    located at the center for a fixed value of the frequency. Right: same for a defect
    located out of plane.
    }
    \label{SW}
\end{center}
\end{figure}

\section{Summary and future.}
Topological defects are likely to be present in the recently
synthesized graphene samples as they are in nanotubes and other fullerenes.
The size and shape observed in \cite{Anetal01} coming from a a single pentagon
is similar to the ripples reported in \cite{Metal06}. This type of disorder
generate long range potentials which have a strong influence on the electronic
properties. In this work we have presented a model for these defects based
on a cosmic striganalogy and the results on the inhomogeneities induced
by this topological disorder on the density of states. This is yet another
example of the fruitful interplay between cosmology and condensed matter\cite{Volovik}
but is has the peculiarity that this time it is cosmology what provides
the model to graphene unlike the usual situation in which condensed matter
systems are used as laboratories to test high energy models.

The effects of the topological disorder on the transport properties
of graphene obtained by averaging over are very interesting
and work in this direction will be reported soon\cite{CV07}.

\section{Acknowledgments} We thank the organizers of the graphene conference
in Dresden for the stimulating atmosphere. We profited there from conversations
on transport in long range correlated disorder
with B. Altshuler, V. Falko, Igor Gornyi, D. Huertas-Hernando and E. McCann.

\end{document}